\documentclass[12pt,preprint]{aastex}
\usepackage{emulateapj5}
\usepackage{apjfonts}
\shorttitle{Multiphase Outflow in FBQS 1044}
\shortauthors{Everett, K\"onigl, \& Arav}

\newcommand{\cc}{\hbox{cm$^{-3}$}}
\newcommand{\persec}{\hbox{s$^{-1}$}}

\begin{document}
\title{Observational Evidence for a Multiphase Outflow in QSO FIRST J1044+3656}
\author{John Everett\altaffilmark{1}, Arieh
K\"onigl\altaffilmark{1,2}, and Nahum Arav\altaffilmark{3,4}}
\altaffiltext{1}{Department of Astronomy and Astrophysics, University
of Chicago, 5640 S. Ellis Avenue, Chicago, IL 60637,
I:everett@jets.uchicago.edu} 
\altaffiltext{2}{Enrico Fermi Institute, University of Chicago,
I:arieh@jets.uchicago.edu} 
\altaffiltext{3}{Astronomy Department, UC Berkeley, Berkeley, CA
94720, I:arav@astron.Berkeley.EDU}
\altaffiltext{4}{Physics Department, University of California, Davis, CA 95616}
\slugcomment{Accepted for publication in ApJ v569 n2, April 20, 2002}

\begin{abstract}
Spectral absorption features in active galactic nuclei (AGNs) have
traditionally been attributed to outflowing photoionized gas located
at a distance of order a parsec from the central continuum
source. However, recent observations of QSO FIRST J104459.6+365605 by
de Kool and coworkers, when intepreted in the context of a
single-phase gas model, imply that the absorption occurs much farther
($\approx 700\ {\rm pc}$) from the center. We reinterpret these
observations in terms of a shielded, multiphase gas, which we
represent as a continuous low-density wind with embedded high-density
clouds.  Our model satisfies all the observational constraints with an
absorbing gas that extends only out to $\sim 4\ {\rm pc}$ from the
central source.
The different density components in this model coexist in the same
region of space and have similar velocities, which makes it possible
to account for the detection in this source of absorption features
that correspond to different ionization parameters but have a similar
velocity structure. This model also implies that only a small fraction
of the gas along the line of sight to the center is outflowing at the
observed speeds and that the clouds are dusty whereas the uniform gas
component is dust free. We suggest that a similar picture may apply to
other sources and discuss additional possible clues to the existence
of multiphase outflows in AGNs.
\end{abstract}


\section{Introduction}
\label{intro}
Modeling active galactic nuclei (AGNs) is a challenging endeavor in
part because of the uncertain geometry of the gas and dust within
their cores.  Our best understanding of the gas distribution comes
from reverberation mappings \citep{BM82,NP97}, which indicate that the
broad emission-line region (BELR) lies at a distance $R_{\rm BELR}
\approx 0.01 L^{1/2}_{44}\ {\rm pc}$ from the central continuum, where
$L_{44}$ is the luminosity in the spectral band $\sim 0.1-1$ {\micron}
in units of $10^{44}$ ergs \persec.  Based on observed absorption of
BELR features by broad absorption-line region (BALR) gas and on
optical polarization measurements, it has been inferred that the BALR
lies outside of the BELR; the best estimates have placed the BALR
somewhere between a few tenths and several parsecs from the central
source \citep[e.g.,][]{T88}.

	Velocities for these outflows are well determined from Doppler
shifts: the BELR and BALR components have speeds $\lesssim 5000~{\rm
to}~8000~{\rm km~s}^{-1}$ and $\lesssim 30000~{\rm km~s}^{-1}$,
respectively.  Other outflow components have also been
detected. Observations of the ``warm absorber'' (a partially ionized
X-ray absorbing gas) in Seyfert galaxies and radio-quiet QSOs, and of
a UV-absorbing component apparently associated with the warm absorber,
indicate gas outflowing at $\lesssim 10^3~{\rm km~s}^{-1}$.
Furthermore, spectral lines classified as being from the Narrow Line
Region (NLR) correspond to velocities of $\lesssim$ several hundred
km~s$^{-1}$.

Different theories have been proposed to explain some of these
outflows.  For example, \citet*{EBS92} considered a
hydromagnetically-driven disk outflow of discrete, long-lived clouds
as a model of the BELR, whereas \citet{M95} and \citet{P00} explored
radiatively-driven, continuous disk winds as the origin of the BELR
and the BALR. These models all involved a single-phase gas medium.

Recently, \citet[hereafter dK01]{dK01} analyzed a Keck HIRES spectrum
of the FIRST Bright Quasar Source J104459.6+365605 (hereafter, FBQS
1044).  They inferred that excited Fe~II levels are not populated
according to local thermodynamic equilibrium. This observation implies
an electron number density $n_{e} \approx 4 \times 10^{3}~{\rm
cm}^{-3}$, which, together with the observed broad absorption in Mg~II
[qualifying this object as a broad absorption-line quasar (BALQSO)],
the Mg~I absorption features, and single gas-phase photoionization
models, place this BALR at $\sim$ 700 parsecs from the central source.
 
Not only does this surprising result contradict the picture of a
close-in BALR, but, in addition, dK01's thorough analysis suggests
that the gas only partially covers the continuum source.  It is not
easy to understand how the absorbing gas could be approximately 1 kpc
from the center and yet only partially cover the AGN core.
Additionally, the Keck spectra indicate distinct groupings of gas in
velocity space; it is also difficult to explain how these apparent
clumps could move to such a large distance and still remain bunched
up, retaining distinct identities.  Even more striking is dK01's
observation that different ionization states have a similar kinematic
structure (i.e., residual intensity as a function of velocity): in
particular, both Fe~II and Mg~I exhibit absorption troughs at -200,
-1250, -3550, and -3800 km~s$^{-1}$.

The $700~{\rm pc}$ distance estimate results from combining the
ionization parameter\footnote{The ionization parameter $U \equiv
Q/4{\pi}nr^2c$ is the dimensionless ratio of the hydrogen-ionizing
photon density to the hydrogen number density $n$: $Q$ is the rate of
incident hydrogen-ionizing photons, $r$ is the distance from the
continuum source, and $c$ is the speed of light.} implied by the
observed low-ionization Mg and Fe absorption lines with the inferred
values of $n_e$ and $L$ (=~$10^{46} {\rm ergs~s^{-1}}$).  One can
reduce this distance estimate by attenuating the incident continuum by
an intermediate gas ``shield.'' In this model, however, ions at
different ionization states (such as the Fe~II and Mg~I components
seen in FBQS 1044) are expected to occupy different regions in space,
with lower values of $U$ corresponding to larger distances from the
center \citep[e.g.,][]{VWK93}. This stratification will likely lead to
disparate velocities for the different ions: for example, models of
self-similar MHD winds and of radiatively-driven constant-$U$ outflows
predict that the terminal velocity, $v_\infty$, varies as $r_{\rm
inject}^{-1/2}$, where $r_{\rm inject}$ is the radius at which the gas
is injected into the outflow \citep*[e.g.,][]{BP82,A94}. Models that
rely solely on shielding and distance to separate the different
ionization components therefore cannot explain dK01's observations, in
which different ionization states are found to have similar
velocities.

We propose to explain the observations of FBQS 1044 by generalizing
the single-phase shielded-gas model, attributing the different
ionization states to different density components in a multiphase
outflow. These components coexist at the same distance from the center
and thus have different ionization parameters but essentially the same
velocities. In particular, if the high- and low-ionization lines
arise, respectively, in a continuous wind and in dense clouds that are
embedded in the outflow, then all the absorption components produced
in a given region of the wind will exhibit similar kinematic
signatures.

In our model, the continuous gas component extends from near the black
hole's event horizon out to the distance where the observed Fe~II
absorption and electron density can be reproduced. The inner part of
this component (interior to the BALR) is identified as the ``shield.''
This region could be associated with an MHD-driven \citep{KK94}, a
Thomson scattering-driven \citep[e.g.,][]{B01}, or a ``failed''
line-driven \citep[e.g.,][]{M95,P00} disk wind, or with a disk corona
\citep[e.g.,][]{EBS92}. The outer region of the continuous gas
component is outflowing and accounts for the Fe~II and Mg~II
absorption, but is still too highly ionized to contain Mg~I absorbing
gas. Within this outflow are embedded higher-density clouds that
account for the lower-ionization Mg~I absorption: their high density
yields a lower U in the clouds, allowing Mg~I to exist.  Such a
two-component outflow may arise naturally in the context of a
centrifugally driven disk wind, which could uplift clouds from the
disk surface by its ram pressure and confine them by its internal
magnetic pressure \citep[e.g.,][]{EBS92,KKE99,EKK01}.  Alternatively,
the clouds may represent transient density enhancements that are
produced by turbulence \citep[e.g.,][]{BF01} or by shocks in a
radiatively driven wind \citep[e.g.,][]{A94}.  For illustration, we
adopt here the ``clouds uplifted and confined by an MHD wind''
picture.

\section{Multiphase Outflow Model}

\subsection{Setup}
Our model consists of two segments that represent two distinct gas
phases.  The first segment, which corresponds to the continuous phase,
extends from $r_{\rm in}=10~GM/c^{2}=7.4 \times 10^{13} \ {\rm cm}$
(following dK01, we take the central black-hole mass to be $M=5 \times
10^{8} M_{\sun}$) out to just past the hydrogen recombination front,
which corresponds to the Fe~III $\rightarrow$ Fe~II transition.  The
second segment, which models the confined dense clouds, is a
constant-density zone located just beyond the continuous gas component
(since the cloud temperature is nearly uniform, this provides a good
representation of magnetically confined constant-pressure clouds). In
reality, the continuous outflow in the line-absorption region can be
expected to engulf the clouds, but this simplified model should
capture the basic physical effects of a two-phase medium.  (We
verified that our results are independent of the cloud radial
distribution within the Fe~II/Mg~I absorption region.)

We obtain the ionization structure of the different gas phases using
the photoionization code Cloudy \citep{F00}, adopting the same source
parameters as in dK01 and assuming (as they do in most of their
models) a \citeauthor{MF87} (1987) spectrum.  We take the scaling of
the hydrogen number density with spherical radius to be $n(r) \propto
r^{-1}$, as in the AGN disk-wind models considered by
\citeauthor{KK94} (1994). In these self-similar MHD outflows, the
magnetic field $B_{\rm wind}$ also scales as $r^{-1}$, which results
in magnetically confined clouds having a constant ionization
parameter.


We first calculate the photoionization of the continuous component,
stopping the computation when $n_{e}$ drops to $4 \times 10^{3}$
cm$^{-3}$ (the value inferred from the observations).  We then
calculate how the radiation emerging through the continuous segment
affects clouds located at that distance.  The two parameters we adjust
in our model are $n_{w,i}$, the wind hydrogen number density at
$r_{\rm in}$, and $n_c$, the hydrogen number density of the clouds. We
explore a range of models to find the values of $n_{w,i}$ and $n_c$
that best reproduce the observations.

\subsection{Results}

Our best-fit value for the wind density is $n_{w,i} \approx 10^{8.75}\
{\rm cm}^{-3}$. This yields the observed $n_e$ in the region (at $r
\approx 4\ {\rm pc}$) where the Fe~II column density in the wind
attains the observed value of $ 3 \times 10^{15}$ cm$^{-2}$, which is
significant since the $n_{e}$ measurement comes from the Fe~II
absorption lines.  However, we do have to cut off the outflow very
close to the end of the hydrogen recombination front so as not to
exceed the observed Fe~II column (see Fig. 1).  Although the abrupt
end of the absorbing column could be an artifact caused by our
simplified treatment, the occurrence of a strong reduction in the
absorption at this location may have an actual physical basis.  It may
be a consequence of the decrease in the ionization fraction in this
region, which reduces the efficiency with which the disk can drive an
MHD outflow, and it may also reflect a transition from a gaseous to a
``clumpy'' disk \citep[see][]{SB87} or from differential to solid-body
disk rotation \citep[see][]{BKS00} on that scale.

We also note from Figure 1 that the Mg~II front occurs $\sim$~2.5
times closer to the central source than the Fe~II front. (This is due
to the formation of a He III $\rightarrow$ He II front in the wind
that absorbs photons with E $>$ 54.4 eV, allowing Mg~III, with an
ionization energy of 80.1 eV, to recombine into Mg~II.)  Since
$v_\infty \propto r^{-1/2}_{\rm inject}$ in our chosen disk-wind model
(see \S \ref{intro}), the predicted smaller initial radius of the Mg
II front is consistent with the higher Mg~II velocities observed in
FBQS 1044.

Besides determining the measured values of the Fe~II column and
of $n_e$, the continuous wind component in our model also exerts a strong
influence on the cloud absorption properties through its effect on the
transmitted continuum spectrum. We find that, in order for the
clouds to account for the observed Mg~I absorption without also dominating
the Fe~II absorption, their iron gas-phase abundance
must be reduced, which we attribute to depletion by dust.  We
assume an ISM dust composition [the default Cloudy graphite and
silicate grains with a \citeauthor{MRN77} (1977; hereafter MRN)
power-law size distribution].  The dust abundance is constrained by
the measured extinction: $A_{2500} \lesssim 1.0$.

Our best-fit value for the typical cloud density is $n_c\approx
10^{8.5}$ cm$^{-3}$.  This means that the clouds are $\sim 10^4$ times
denser (and have a correspondingly lower ionization parameter) than
the surrounding continuous wind.  Figure 2 shows the absorption column
densities associated with a string of such clouds as a function of
pathlength through the clouds.

Our model can account for the observations with cloud densities that
range between $10^{7.75}$ and $10^{9}$ cm$^{-3}$. This leeway reflects
the uncertainty in $A_{2500}$. Whereas dK01 employ an extinction of
1.0 magnitude, $A_{2500}$ could range from $\sim 1.0$ down to 0.2
magnitudes based on a comparison between the spectrum of FBQS 1044 and
that of an ``average'' QSO \citep{W91}. If a population of
lower-density clouds were present with a significant fraction of the
observed Mg~I column, then the dusty clouds would have $A_{2500} > 1$.
On the other hand, if $n_c \gg 10^{9}$ cm$^{-3}$, $A_{2500}$ would
drop below 0.25.  With $n_c = 10^{8.5}$ cm$^{-3}$ we predict $A_{2500}
\approx$ 0.5 magnitudes, which fits well within the inferred range.

With the extinction already accounted for, the data require the
continuous gas component to be effectively dust free. We tested this
hypothesis by including Cloudy's Orion-type dust (a set of graphite
and silicate grains with MRN's minimum size increased from 0.0025 to
0.03 {\micron}, appropriate to a UV-irradiated medium) in the wind,
taking into account dust sublimation and sputtering.  To include the
effects of the latter process, we removed all dust grains whose
sputtering time scales (calculated following \citealt{T94}) do not
exceed 1000 yr (the approximate time grains spend in the wind if they
travel at the observed outflow velocity over a distance of $1\ {\rm
pc}$).  Including both of these effects, the predicted dust extinction
is over one order of magnitude greater than the maximum value allowed
by the observations.  We also considered the effect of changing the
scaling of the wind density with radius from $n \propto r^{-1}$ to $n
\propto r^{-3/2}$ (as in \citealt{BP82}) and to $n \propto
r^{-2}$. These models again overestimate the extinction by about an
order of magnitude after dust sublimation and sputtering are taken
into account.  We therefore conclude that, if the gas comprising the
shield is associated with a disk outflow, then it must already be dust
free when it leaves the disk. Such a situation could arise if the wind
originates in a hot disk corona where the matter resides long enough
for any dust grains to be destroyed.

However, to provide the requisite shielding, a dust-free wind requires
a higher total gas column than a dusty outflow, so the inferred lack
of dust in the wind implies a prohibitively large mass outflow rate
--- much larger than could be launched by the magnetic field that
confines the clouds in our model.  Thus, only part of the shield can
be outflowing at velocities that are comparable to (or exceed) the
observed speeds.  Perhaps the inner outflow has a lower velocity than
we predict, or else some of the shield may not even be outflowing, as
in a disk corona or a ``failed'' line-driven wind (see \S
\ref{intro}).

All of these considerations yield our best model, which is compared
with the observational results in Table 1.  This model satisfies all
the observational constraints and implies that the absorbing gas lies
over two orders of magnitude closer to the central source than the
earlier estimate.  The radius where the observed gas leaves the disk
surface is, in general, smaller yet.

We can estimate the mass outflow rate associated with the absorbing
gas through the relation $\dot{M}_{\rm wind} \approx 2 \pi r f N_{\rm
H} m_{p} v$, where $f$ is the fraction of $4 \pi$ steradians into
which the wind flows, $N_{\rm H}$ is the total hydrogen column density
of {\em only}\/ the inferred Mg~II and Fe~II absorbing region
($\approx 3.9\times 10^{23}$ cm$^{-2}$), $v$ is the observed outflow
speed ($\sim 10^{8}$ cm s$^{-1}$), $r$ ($= 1.2 \times 10^{19}$ cm) is
the inferred distance, and where we assume a vertically and
azimuthally continuous wind.  We take $f \sim 0.1$ for BALQSO sources
\citep{W97}.  For the above values, we find $\dot{M}_{\rm wind}
\approx 8$ $M_{\odot}$ yr$^{-1}$.

By equating the thermal pressure in the clouds ($\approx 2.7 \times
10^{-4}$ dynes cm$^{-2}$) to the confining wind magnetic pressure,
$B^{2}_{\rm wind}/8 \pi$, we deduce $B_{\rm wind} \approx 8.2 \times
10^{-2}\ {\rm G}$. It is encouraging that, when this value of $B_{\rm
wind}$ is used in the $n(r) \propto r^{-1}$ self-similar MHD wind
model, it implies a local mass outflow rate that is comparable to the
above estimate of $\dot{M}_{\rm wind}$, yielding $\sim 4\ M_{\odot}\
{\rm yr}^{-1}$.  If we instead choose $n_c = 10^{9}$ cm$^{-3}$ (so as
to satisfy the lower limit on the dust extinction, $A_{2500} = 0.25$)
and require pressure balance, we find $\dot{M}_{\rm wind} \approx 9\
M_{\odot}\ {\rm yr}^{-1}$.

\section{Conclusion}

The most robust result of our study is that a shielded, multiphase
absorption region reproduces the observations of FBQS 1044 on a
conventional BALR scale ($\approx 4$ pc).  In addition, when one
attributes the Fe~II and Mg~II absorption to a low-density outflow
component and the Mg~I absorption to a cospatial high-density outflow
component, it is possible to explain the similar kinematic structure
of the respective spectral features.  We also find that only a small
fraction of the gas along the line of sight can be outflowing at the
observed speeds and that
only the high-density component of the
outflow is dusty.  We derived these results using a ``clouds embedded
in a continuous MHD disk wind'' model, but our conclusions also apply
to other plausible scenarios that include a continuum-shielding gas
column and an absorption region that contains distinct low- and
high-density components. Our basic conclusions appear to be quite
general, although the precise composition of the absorbing gas and its
detailed spatial and kinematic properties are not fully constrained by
the observations and remain model dependent.

In addition to explaining the FBQS 1044 observations, this picture may
be relevant to the interpretation of absorption features in similar
objects where single-phase models imply a large distance.  For
instance, in the case of the radio-loud galaxy 3C 191, absorber
distances of $\sim 28$ kpc were inferred by \citet{H01} using similar
arguments to those employed by dK01.  A multiple-phase model could
place these absorbers much closer to the central source.

This interpretation may also be applicable to other AGN observations.
As outlined in \S~1, several distinct outflow components have been
inferred in various types of AGNs. There is now growing evidence that
these components may not be single-phase. For the warm-absorber
component, which has been inferred to give rise to both X-ray and UV
absorption \citep[e.g.,][]{C97,MWE98,MMWE01},
there are indications in at least some sources that the X-ray and UV
absorbing components are not identical (e.g., the Seyfert 1 galaxy NGC
3783 --- \citealt*{K00, KCG01}).  Furthermore, it appears that the UV
and X-ray absorbing components are themselves divided into multiple
zones. In the UV regime, this has been established in Seyfert 1
galaxies like NGC 3783 \citep{KCG01} and NGC 3227 \citep{CKBR01}.  In
the X-ray regime, this is exemplified by the Seyfert~1 galaxy
MCG-6-30-15, which was modeled by \citet*{MFR00} as an extended
multi-zone medium. These authors suggested that, in reality, these
zones may correspond to a continuum of clouds at different radii and
different densities, as in the BELR model of \citet{B95}. A
hierarchical, turbulent-gas realization of this concept was recently
presented by \citet{BF01}. Our ``clouds in a continuous wind''
scenario is designed to explicitly address the outflowing nature of
the absorbing gas but is otherwise similar to the above picture in its
description of spatially coexisting multiple phases. For simplicity,
we have treated the clouds as long-lived, pressure-confined entities,
but it is entirely conceivable that the clouds are transient features
that arise in a turbulent outflow. The same basic picture of a
multiphase medium may thus apply to the gas in the BELR, the BALR
\citep[e.g.,][]{A99}, the warm absorber, and even the NLR
\citep[e.g.,][]{K01}.

In conclusion, we have argued that the observations of FBQS 1044 can
be interpreted in the context of the standard BALR picture in terms of
a gas outflow that consists of (at least) two phases, which we modeled
as a continuous wind with embedded dense clouds, shielded from the
central continuum.  If the shield is identified with the continuous
component, then only a fraction of it can be outflowing at a speed
that approaches (or exceeds) the value in the Fe/Mg absorption
region. We also deduced that the clouds are dusty but that the shield
is effectively dust free.  As far as we are aware, this is the first
instance of a BALR outflow in which the data provide direct evidence
for the existence of a multiphase medium. Together with other pieces
of evidence, this result lends support to the view that all the major
outflow components in AGNs may contain multiple phases.

\acknowledgements{We gratefully acknowledge the initial work of and
continued discussions with Martijn de Kool as well as helpful comments
from Mike Brotherton, Abraham Loeb, and the referee.  J.E. and
A.K. thank NASA for support through grant NAG5-9063. }

\begin{table}
\centering
\caption{Comparison of the Two-Phase Outflow Model with Observational
Results for FIRST QSO 1044+3656
\label{tbl-1}}
\normalsize
\begin{tabular}{|l|c|c|}
\hline
Quantity           &  Observational Result              &  Model Result  \\
\hline \hline
N(Fe~II)           & $\sim 3 \times 10^{15} $cm$^{-2}$  &  $3 \times 10^{15} $cm$^{-2}$ \\       
N(Fe~I)            & $< 10^{13} $cm$^{-2}$              &  $2.5\times 10^{11} $cm$^{-2}$  \\
N(Mg~II)/N(Mg~I)   & $\gtrsim 30$    &  500 \\
N(Mg~I)            & $\sim 2 \times 10^{13} $cm$^{-2}$  &  $2 \times 10^{13} $cm$^{-2}$ \\
$n_{e}$            & $\sim$ 4000                        &  4000 \\
A$_{2500}$         & $\lesssim$ 1     &  $\sim$ 0.5 \\
\hline \hline
\end{tabular}
\end{table}

\clearpage
\begin{figure}
\figurenum{1} 
\plotone{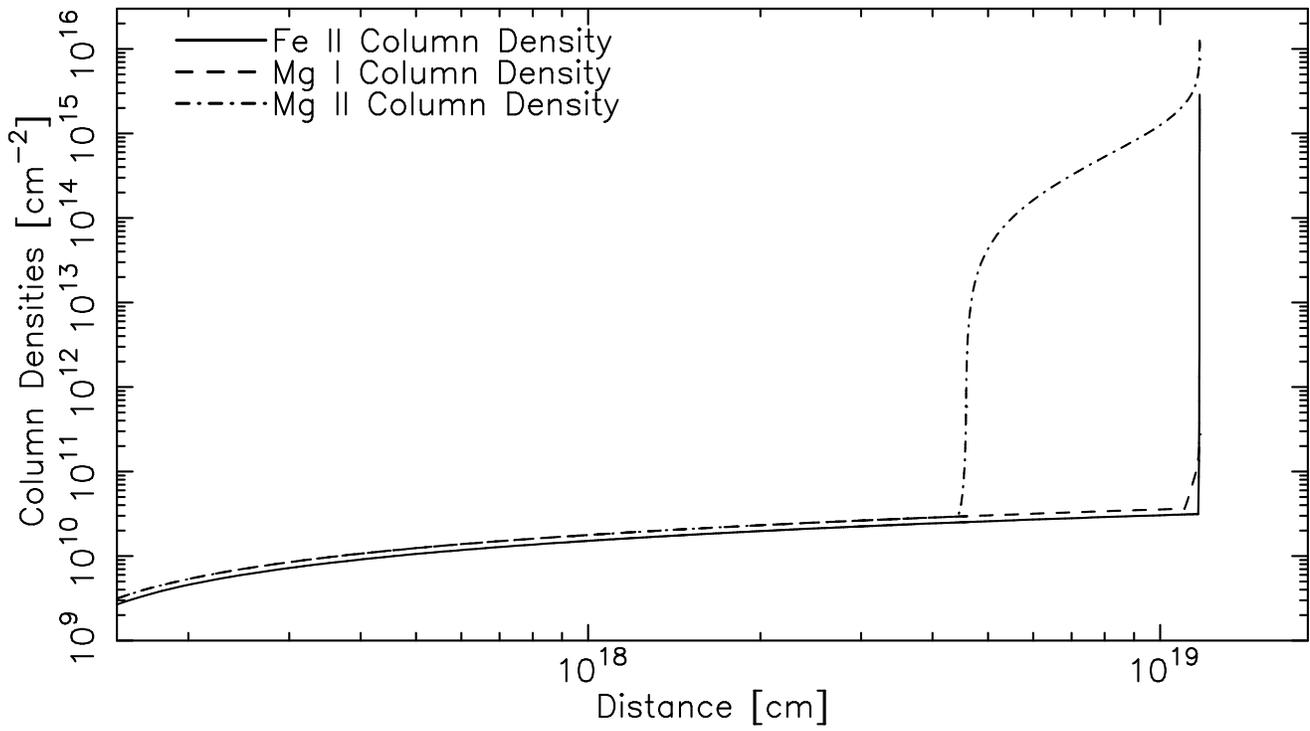} 
\figcaption{Column densities in the dust-free continuous wind model
parameterized by $n_{w,i} \approx 10^{8.75}\ \cc$.  The continuous
wind accounts for the Fe~II as well as the higher-velocity Mg~II
absorption without overproducing the low-ionization Mg~I absorption
(which is attributed to the dense clouds).}
\end{figure}

\clearpage
\begin{figure}
\figurenum{2}
\plotone{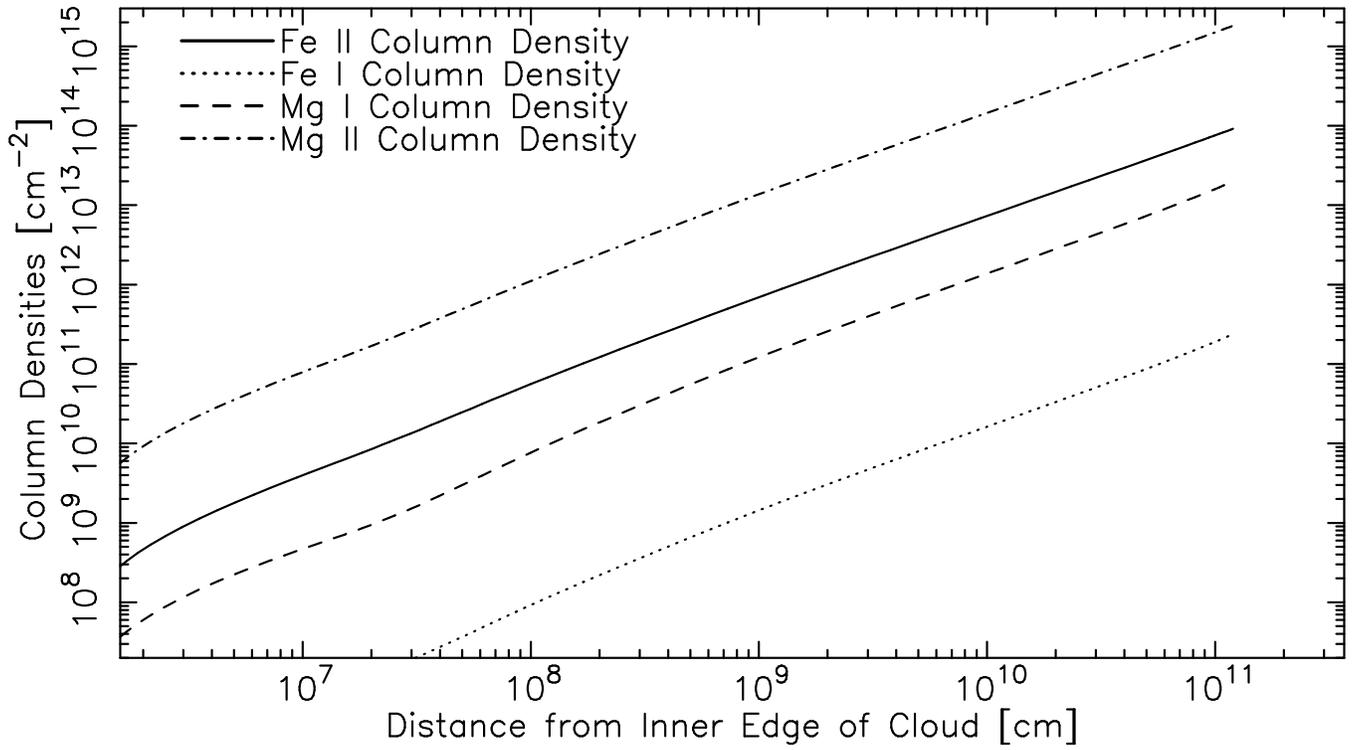}
\figcaption{Column densities in dusty clouds of hydrogen density
$n_c=10^{8.5}\ \cc$ that are shielded by the continuous dust-free wind
of Fig. 1.  The clouds account for the Mg~I absorption without
overproducing the higher-ionization Fe~II absorption (which is
attributed to the continuous wind).}
\end{figure}

\end{document}